\documentclass[authoryear,preprint,review,12pt]{elsarticle}

\usepackage{amssymb}
\usepackage{lipsum}

\usepackage{hyperref}

\usepackage[final]{changes}

\journal{Icarus}

\begin{document}

\begin{frontmatter}



\title{Impacting the dayside Martian ionosphere from above and below: Effects of the impact of  CIRs and ICMEs close to aphelion (April 2021) and during  dust storms (June/July 2022) seen with MAVEN ROSE}


\author[first]{Marianna Felici}
\affiliation[first]{organization={Center for Space Physics, Boston University},
            addressline={725 Commonwealth Avenue}, 
            city={Boston},
            state={MA},
            country={USA}}

\author[second]{Jennifer Segale}
\affiliation[second]{organization={Department of Astronomy, Boston University},
            city={Boston},
            state={MA},
            country={USA}}
            
\author[third]{Paul Withers}
\affiliation[third]{organization={Department of Astronomy, Boston University},
            city={Boston},
            state={MA},
            country={USA}}

\author[forth]{Christina O. Lee}
\affiliation[forth]{organization={Space Science Laboratory ,University of California},
            city={Berkeley},
            state={CA},
            country={USA}}

\author[fifth]{Andrea Hughes}
\affiliation[fifth]{organization={NASA Goddard Space Flight Center},
            city={Greenbelt},
            state={MD},
            country={USA}}
            
\author[sixth]{Ed Thiemann}
\affiliation[sixth]{organization={LASP, University of Colorado},
            city={Boulder},
            state={BO},
            country={USA}}

\author[seventh]{Steve Bougher}
\affiliation[seventh]{organization={University of Michigan},
            city={Ann Arbor},
            state={MI},
            country={USA}}

\author[eight]{Candace Grey}
\affiliation[eight]{organization={NMSU Astronomy Department},
            city={Las Cruces},
            state={NM},
            country={USA}}
            
\author[ninth]{Shannon Curry}
\affiliation[ninth]{organization={LASP, University of Colorado},
            city={Boulder},
            state={CO},
            country={USA}}


\begin{abstract}
We use 62 electron density profiles collected by the Radio Occultation Science Experiment (ROSE), on the Mars Atmosphere and Volatile EvolutioN (MAVEN), when Mars was hit by CIRs and ICMEs close to aphelion (April 2021) and during two dust storms (June/July 2022) to examine the response of the Martian ionosphere to solar events and to solar events hitting during dust storms. We do so through three proxies – variation in total electron content between 80 and 300 km altitude, peak density, and peak altitude – of the aforementioned 62 ROSE electron density profiles, relative to a characterisation of the ionosphere through solar minimum leading to solar maximum, specific to local time sector and season, presented in the \href{https://arxiv.org/abs/2312.00734}{https://arxiv.org/abs/2312.00734} by Segale et al., [\href{https://arxiv.org/abs/2312.00734}{https://arxiv.org/abs/2312.00734}].
We observe an increased Total Electron Content (TEC) between 80 and 300 km altitude up to $\simeq$ 2.5 $\times 10^{15}$ m$^{-2}$ in April 2021 and up to $\simeq$ 5 $\times 10^{15}$ m$^{-2}$ in June/July 2022 compared to the baseline photochemically produced ionosphere. This increase in TEC corresponds mainly to increases in the solar energetic particles flux (detected by MAVEN Solar Energetic Particle, SEP) and electron fluxes (detected by the MAVEN Solar Wind Electron Analyzer, SWEA). In addition to solar events, in June/July 2022, an A storm and a B storm were occurring and merging on the surface of Mars. We  observe a raise in peak altitude in general lower than expected during dust storms, possibly due to high values of solar wind dynamic pressure (derived from MAVEN Solar Wind Ion Analyzer, SWIA).  From 31 ROSE profiles collected in this time period that showed both the M2 and M1 layer, we observe that, on average, M1 and M2 peak altitudes raise the same amount, suggesting that the thermosphere might loft as a unit during dust storms. During this time period, several proton aurora events of variable brightness were detected with MAVEN Imaging Ultraviolet Spectrograph (IUVS), underlining the complex and multifaceted impact of dust activity and extreme solar activity on the Martian ionosphere.
\end{abstract}


\begin{highlights}
\item \added{Increase in TEC up to $2.5 \times 10^{15}$ m$^{-2}$ in the Martian ionosphere between 80 and 300 km following an ICME/CIR impact.}
\item \added{Increase in TEC up to $5 \times 10^{15}$ m$^{-2}$ in the Martian ionosphere between 80 and 300 km following ICME/CIR impacts during dust storms.}
\item \added{During dust storms, the peak altitudes of the Martian ionospheric layers M1 and M2 loft of the same amount.}
\end{highlights}


\end{frontmatter}




\section{Introduction}
\label{introduction}

At Mars, the ionisation in the dayside ionosphere, a region of the atmosphere embedded in the thermosphere, is mainly controlled by photochemical processes such as CO$_{2}$ + photon $\rightarrow$ CO$_{2}^{+}$ + $e$, CO$_{2}^{+}$ + O $\rightarrow$ O$_{2}^{+}$ + CO, O$_{2}^{+}$ + $e$ $\rightarrow$ O + O) below about 170-200 km. At those altitudes, in fact, photochemical processes act faster than plasma transport \cite[e.g.][]{Barth:1992aa,Fox:2004aa, Withers:2009aa,Bougher:2017} \replaced{and}{:} the behaviour of the peak densities as a function of solar flux and of solar zenith angle can generally be represented quite well by the idealised photochemical theory  for both the M2 layer (created by EUV radiation) and M1 layer (created mainly by X-rays radiation and secondary electrons) \citep{Chamb:1987, Withers:2009aa}.

Mars is immersed in the heliosphere and, therefore, is subject to space weather and solar events. 
For example, the interaction between the stream of faster solar wind originating from solar coronal holes and the stream of preceding slower solar wind forms a region of compressed plasma, the stream interaction region (SIR). If the SIR, twisted into a spiral by the rotation of the Sun, persists for more than one solar rotation, then it is referred to as Corotating Interaction Region (CIR)  \cite[][and references therein]{Richardson:2018aa}. The in situ signatures associated with SIRs and CIRs can be enhancements in the solar wind dynamic pressure; solar wind particle speed and density; frequent pure energetic ion enhancements, namely no enhancement in the flux of energetic solar wind electrons, but only in the flux of ion with energies $\sim$ MeV$/$n); less frequent pure electron enhancements ($\sim$ 0.04 - 1 MeV); enhancements in both energetic ion and electron; and increase in solar wind magnetic field strength \cite[][and references therein]{Richardson:2018aa}. 

Another example of solar events are coronal mass ejections (CMEs) which, instead, are bubbles made of billions of tons of solar plasma, with embedded magnetic fields, that erupt from the Sun and travel outwards in the Interplanetary heliosphere (ICME) at a pace between 250 and 3000 km$/$s \cite[][and references therein]{Jones:2020aa}. 
There are a plethora of in situ signatures in the interplanetary medium associated with a CME, depending on whether the CME belongs to the magnetic cloud or non-cloud events categories, and if it is a fast or slow event \citep{Owens:2018aa}.
These signatures do not appear necessarily at the same time, nor in the same region of the heliosphere \citep{Zurbuchen:2006aa, Owens:2018aa}; some of these, at 1 AU, can be rotation and enhancement of the magnetic field, forward shock, increase in the $\alpha$ to protons ratio, high energy particles  \citep{Zurbuchen:2006aa}. The latter, Solar Energetic Particles (SEPs) (energies in the keV–GeV) can increase in flux: these events can last from hours to days in a large range of heliolongitudes from the Sun \citep{Palmerio:2021aa} and be associated also to flare events.

Previous studies reported the response of the Martian atmosphere to such events mostly focusing on either the top or the bottom of the ionosphere.

Regarding the effects of CIRs, \cite{Dubinin:2009aa} showed that the impact of a dense and high pressure solar wind can penetrate the Martian induced magnetosphere and induce erosion channels in the ionosphere above 300 km. \cite{Edberg:2010aa}, in a statistical study that confirms the findings of previous case studies \cite[e.g.,][]{Dubinin:2009aa}, finds that pressure pulses, whether from CIRs or ICMEs, lead to an increase in ions lost from the planet. \cite{Ram:2023aa} suggest that the impact of CIR on the Martian ionosphere is larger compared to CME impacts in a declining phase of the solar cycle.


\cite{Ulusen:2012aa} studied the effect of multiple Solar Energetic Particles (SEP) events impacting the ionosphere of Mars and found that in some instances the electron density between 100 and 200 km altitudes did not show any increase, however, some of those events did show an increase in density below 100 km altitude, suggesting that superthermal electrons (10$-$20 keV) could increase ionization below the M1 layer (lower ionosphere); fewer events at limited solar zenith angles, instead, show lower electron density between 100 and 120 km altitude or even at the top of the ionosphere, suggesting that SEP increase atmospheric loss and ionospheric compression. \cite{SanchezCano:2019aa} found solar energetic electron precipitation to be responsible for creating a low ionospheric plasma layer at $\sim$90 km altitude on the Martian nightside, layer that might have cause days long blackouts in the Mars Advanced Radar for Subsurface and Ionosphere Sounding (MARSIS) onboard Mars Express and the Shallow Radar (SHARAD) onboard the Mars Reconnaissance Orbiter. \cite{Lester:2022aa} found clear correlation between the blackouts of MARSIS and SHARAD instruments and solar cycle, and estimated that, to cause blackout in both instruments, the average SEP energy spectrum is above 70 keV.

\cite{Opgenoorth:2013aa} found that increases in the dynamic pressure of the solar wind would trigger compression in the magnetosphere and ionosphere and, as a consequence, increase plasma transport over the terminator and ion escape. \cite{Morgan:2014aa}, also, observed that the ionosphere after an ICME impacting Mars extends well beyond the terminator, up to solar zenith angle of 115$^{\circ}$, with a few possible causes such as the increased plasma transport from the dayside, electron precipitation, and increased solar energetic particles precipitation.


The effect of an ICME on the upper atmosphere was also reported by \cite{Jakosky:2015CME}, who observed changes in bow shock and magnetosheath, aurora formation, and enhancement in the pick-up ions population with MAVEN instruments. 

From all events that have an impact on the Martian ionosphere from below, we focus on this work on dust \cite{Leovy:2001aa, Kahre:2017aa}, that can have dramatic effects in elevating the altitude of the main ionospheric peaks more than $\sim 15$ km even if the dust storm is not global to the whole planet \citep{felici2020}. 
\cite{Withers:2018aa}, in an earlier analysis of MAVEN ROSE observation during a small dust event, found that the dayside ionospheric peak altitudes were enhanced by more than $\sim$ 9 km at 52$^{\circ}$N in response to said small dust event.
As far as previous large radio occultation datasets, Mars
Global Surveyor (MGS) acquired 5600 ionospheric electron density profiles, but these observation were confined at high northern latitudes ($>$ 60$^{\circ}$ N) and had unfortunate seasonal biases in their coverage. However, \cite{withers2013} identified an instance in Mars Year (MY) 27 where ionospheric peak altitudes observed at  70$^{\circ}$N increased by a highly-uncertain 5 km, coincident with a small increase of the dust content of the tropical atmosphere.

\cite{Girazian:2019} studied dust events during six different Mars Years (MY) using data from the MARSIS (Mars Advanced Radar for Surface and Ionosphere Sounding) instrument on Mars Express. They report an increase $\sim$10--15 km during all six event, regardless of whether these were local, regional, or global storms.




The aim of this article is to utilise the baseline for the ionosphere through solar minimum leading to solar maximum provided by the work of Segale et al., [\href{https://arxiv.org/abs/2312.00734}{https://arxiv.org/abs/2312.00734}] to separate  solar events and dust effects on the Martian ionosphere from the baseline photochemically produced Martian ionosphere. We do this for altitudes between 80 and 300 km, when the Martian ionosphere is impacted by CIRs and ICMEs, during aphelion season, defined here as the ranges 0-161$^{\circ}$ and 341-360$^{\circ}$ L$_{S}$, non-dust season\deleted{)}; and perihelion season, defined here as the range 161-341$^{\circ}$ L$_{S}$\deleted{)}, which corresponds to dust season,  during two merging dust storms. To achieve our aim, we utilise electron density profiles from the MAVEN Radio Occultation Science Experiment (ROSE) \citep{Withers:2020um}, from which we define three proxies: variation from the baseline ionosphere of total electron content, peak density, and peak altitude.


The structure of this article is as follows:
Section \ref{methods} describes the observations analysed in this work and the methodologies with which these were analysed.
Section \ref{results} reports and interprets MAVEN ROSE observations obtained during ICMEs and CIRs impacts collected in 2021 and 2022.
Section \ref{conclusion} summarises the findings of this work.

\section{Methods}
\label{methods}
The Radio Occultation Science Experiment (ROSE) \citep{Withers:2020um} has been part of the scientific payload of the Mars Atmosphere Volatile EvolutioN (MAVEN) spacecraft \citep{Jakosky:2015} since 2016. ROSE conducts two-way X-band radio occultations, providing seasonal observations of the vertical structure of the ionosphere, including the main plasma layer and below. That region of the ionosphere ($\sim$ 130 km and below) has rarely been accessible to other in situ instruments on MAVEN. ROSE has collected more than 1000 electron densities profiles since 2016, with global geographical coverage ($\sim$ -85 to 89$^{\circ}$ in latitude and $\sim$ -180 to 180$^{\circ}$ in longitude) and broad Solar Zenith Angle (SZA) coverage ($\sim$ 47 to 136$^{\circ}$).

To evaluate the effects that ICMEs and CIRs impacts have on the ionosphere of Mars, and separate them from the baseline photochemical ionosphere, between 80 and 300 km altitude, both when the ionosphere is otherwise undisturbed and when it is, instead, heated by the presence of dust and additionally lofted from below by dust storm, we restricted this study to time intervals for which we had a series of ROSE electron density profiles that broadly covered the dayside ionosphere (SZA $< 95^{\circ}$).

\begin{figure}
	\centering 
	\includegraphics[width=\textwidth]{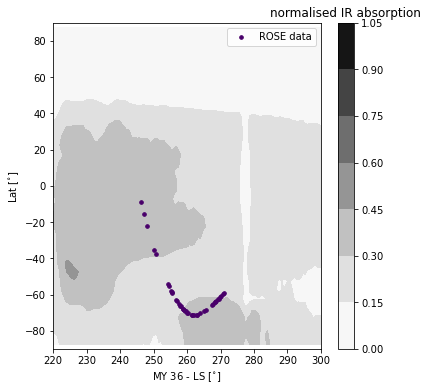}	
	\caption{June/July 2022: ROSE data coverage overlapped with a dust map obtained from the Mars Climate Database \citep{Montabone:2015aa, Montabone:2020aa}.} 
	\label{fig_1}%
\end{figure}

We selected two time periods during which the ionosphere of Mars is impacted by CIRs and/or ICMEs: 
\begin{enumerate}
    \item {\bf April 2021}: we utilised 21 ROSE electron density profiles collected between April 9th and April 23rd 2021. These profiles cover a range in SZA between $\simeq$ 71 and 94$^{\circ}$. As for Local Solar Time (LST), all the profiles were collected consistently in the afternoon at $\simeq$ 16:00, and, latitudinally, in the southern hemisphere; the longitudinal sampling is homogeneous. L$_{S}$ in this time interval ranges between 29 and 36$^{\circ}$, therefore we are close to aphelion (defining aphelion here as the ranges in  L$_{S}$ 0-161$^{\circ}$ and 341-360$^{\circ}$). While we do have MAVEN in situ measurements for this time period, MAVEN was not upstream from Mars, therefore determining the exact ICME and CIR arrival times is not possible. However, \cite{Dresing:2023aa} report  about a long-lasting solar eruption on 17 April 2021 which generated a Solar Energetic Particle (SEP) event so widespread that was detected by different observers in the heliosphere (see \cite{Dresing:2023aa}, Figure 14, for Mars). We report MAVEN Solar Wind Electron Analyzer (SWEA \citep{mitchell2016})  and Solar Energetic Particle (SEP) data  \citep{larson2015} in Figure \ref{fig_a1}, and in Figure \ref{fig_3}: we see a first peak in particle flux around April 12th-13th, one around the 19th, and another, larger, on the 22nd.

    \item  {\bf June/July 2022}: We utilised 41 ROSE electron density profiles collected between June 13th and July 22th 2022. These profiles cover a SZA range between $\simeq$ 47 and 95$^\circ$ - very broad coverage and the lowest in SZA ROSE had ever gotten - spanning nearly the entire dayside range permitted by radio occultation geometry. As for LST, most of these profiles span from $\sim$ 01:00 until 09:00 LST, therefore mostly dawn sector/noon, and 11 of these were collected at dusk after $\sim$ 2300.  Latitudinally, all profiles were collected in the southern hemisphere (see Figure \ref{fig_1}). L$_{S}$ spanned in this time interval ranges between 246 and 271$^{\circ}$, therefore we are at perihelion, in full dust season. Not only that, but, at this time, there were two dust storm merging at Mars, below the ionosphere: an A class storm ('a classic regional-scale or planet encircling Southern Hemisphere dust event' \cite{Kass:2016aa}) which started on May 2nd, at $\sim$ 219 $L_{S}$, peaked at $\sim$ 227 $L_{S}$, and ended at $\sim$ 259 $L_{S}$; and a B class storm ('southern polar event' \cite{Kass:2016aa}), which started  on June 28th,  at $\sim$ 255 $L_{S}$, peaked at $\sim$ 265 $L_{S}$, and ended at $\sim$ 298 $L_{S}$.
    In Figure \ref{fig_1} we show ROSE data coverage in this time period, overlapped with the normalised IR absorption \citep{Montabone:2015aa, Montabone:2020aa}: \deleted{besides the only profile collected at northern latitudes, }ROSE data crossed the two storms fully. Between $\sim$250 and 260$^{\circ}$  L$_{S}$ some ROSE data is located at latitudes outside the dustiest areas (see Figure \ref{fig_1}); however, the timescale with which the thermosphere recovers from dust effects can range between 20 and 120$^{\circ}$ L$_{S}$ \citep{withers2013}, therefore these profiles can still be considered affected by dust.
While we do have MAVEN in situ measurements for this time period as well (see Figure \ref{fig_a2} and \ref{fig_6}), MAVEN did not sample the upstream solar wind conditions until July 6th 2022 (see Figure \ref{fig_a3}, \citep{halekas2015, Halekas:2017aa, Halekas:2015aa}), bottom three panels), therefore determining the exact ICMEs/CIRs arrival time is not possible. However, we do see a large increase in SEP fluxes starting around July 12th and culminating on July 13th/14th, and smaller ones at the end of June/beginning of July.
\end{enumerate}

We use the simulation results of the modelled ICME propagation and ambient solar wind conditions produced from the ENLIL heliospheric solar wind model \citep{Odstrcil:2003aa}, given that during both these time periods 
Earth and Mars were located at $\leq 90^{\circ}$ from each other on the ecliptic plane (see Figure \ref{fig_a1} and \ref{fig_a2}).
The model has limitation and uncertainties ($\sim$ 24 hours at Mars for both shock arrival time and ejecta arrival time \citep{Palmerio:2021aa}), for example given by the choice of input of the CME injection direction, time, size,  and shape  \citep[][and references therein]{Lee:2018aa}, however, ENLIL  has been found to be a good first order approximation of the heliospheric conditions associated with the CME \citep[e.g.][]{Palmerio:2021aa}. 

For this study, we use the ENLIL simulation results for obtaining a more global context of interplanetary conditions at Mars, and, as a tool for interpreting the in situ MAVEN measurements during space weather activity, particularly when the spacecraft was not located upstream, i.e., outside the Martian bow shock, to directly observe the SIR/CIR or ICME impacts \citep[e.g.,][and references therein]{Lee:2018aa}.  The ENLIL simulation results are publicly available through the NASA Community Coordinated Modeling Center (CCMC) Space Weather Database Of Notifications, Knowledge, Information (DONKI) website \url{https://kauai.ccmc.gsfc.nasa.gov/DONKI/search/}.

We report in Figures \ref{fig_a1} and \ref{fig_a2} particles and field data and ENLIL images for these two time periods.

Based on the ENLIL simulation results, in April 2021, the first increase in particle flux might be due to a CIR impacting Mars around April 11th, the second to a glancing blow from a ICME impacting Mars on April 18th, the third to a direct ICME impact on Mars on April 22nd (Figure \ref{fig_a1}).

In June/July 2022, based on the ENLIL simulation results, there may have been an ICME impact at Mars on June 18th and a glancing blow ICME impact combined with a CIR on June 24th, and ICME with glancing blow on July 2nd, a ICME on July 13th, and a glancing blow from a ICME combined to a CIR on July 14th (Figure \ref{fig_a2}). 




We identified three quantities that could help us evaluate the effects of these different drivers on the ionosphere and quantify the deviation that these induce from the undisturbed photochemically produced ionosphere: the Total Electron Content (TEC) between 80 and 300 km altitudes (as above 300 km the plasma population could instead be depleted by the Solar Wind (SW) \citep{Dubinin:2009aa}); the peak electron density, which normally in the dayside coincides with the M2 peak density; and the peak altitude, namely the altitude at which we have highest electron density, which, normally in the dayside tends to coincide with the M2 peak altitude. 

\begin{figure}
	\centering 
	\includegraphics[width=1\textwidth]{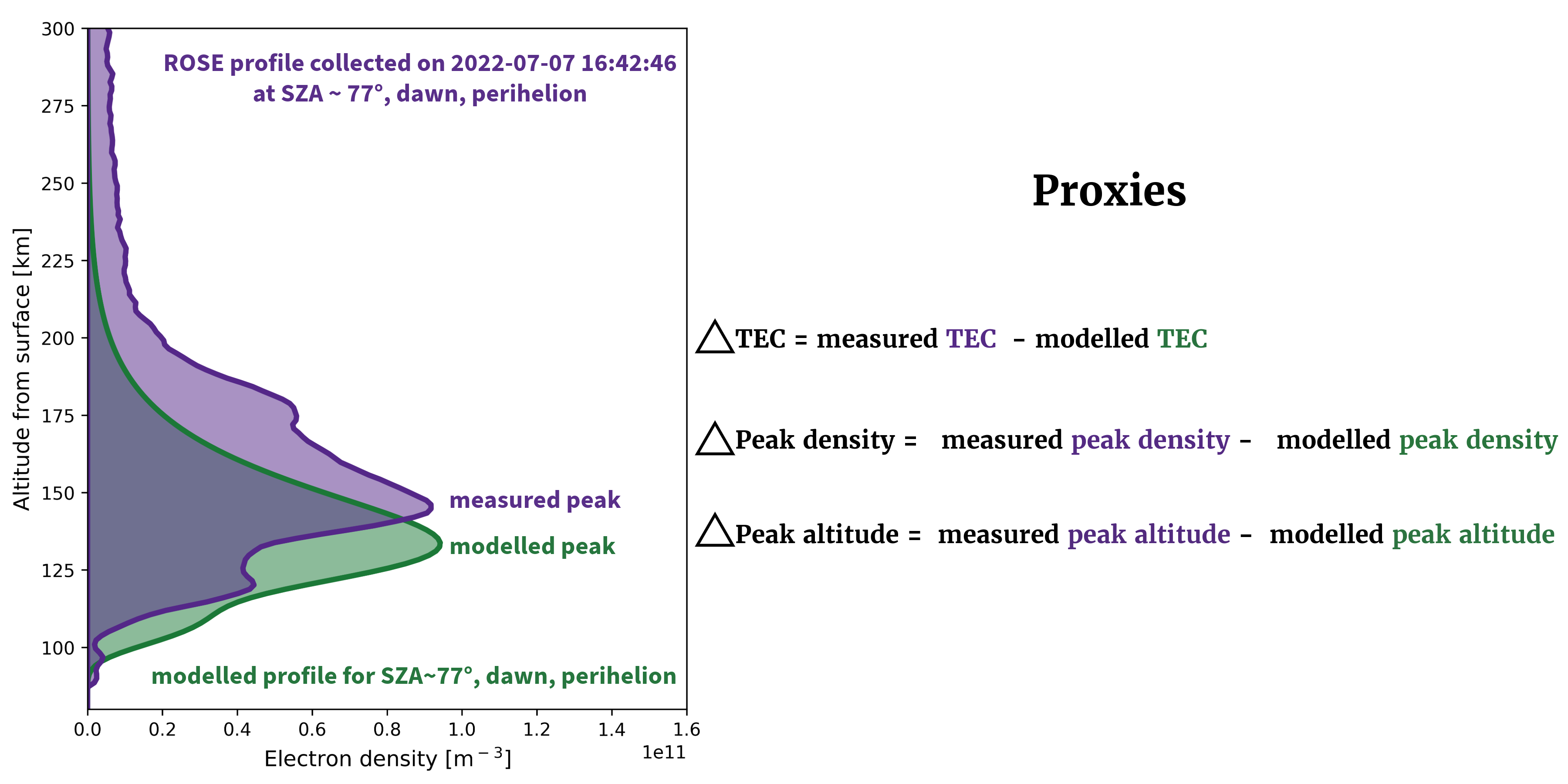}	
	\caption{Example of a measured ROSE electron density profile (left, in purple), during July 2022, when the ionosphere of Mars was hit by  drivers CIR, ICME, and dust storm. 
 We report the corresponding modelled electron density profile - for a merely photochemically produced undisturbed ionosphere [Segale et al., \href{https://arxiv.org/abs/2312.00734}{https://arxiv.org/abs/2312.00734}] - with same SZA, LST sector, and season (left, in green). The  deviation of the measured profile from a photochemical ionosphere is visible. The three proxies derived from the TEC, peak altitude, and peak densities for sets of profiles A and B are reported on the right: these will will help us quantify the deviation induced by drivers from the undisturbed photochemically produced ionosphere.} 
	\label{fig_2}%
\end{figure}

To quantify the effects that space weather events and dust storm induce on the ionosphere, we need to isolate and subtract the baseline photochemically produced ionosphere first. In order to do this accurately, our baseline photochemically produced ionosphere must account for changes that solar local time and season induce on the ionosphere, because these also play a role [Segale et al., \href{https://arxiv.org/abs/2312.00734}{https://arxiv.org/abs/2312.00734}].
We report here summary and conclusions from Segale et al., [\href{https://arxiv.org/abs/2312.00734}{https://arxiv.org/abs/2312.00734}] \emph{"219 electron density profiles of the Martian undisturbed dayside ionosphere collected by MAVEN ROSE between July 2016 and December 2022 through solar minimum leading to solar maximum, show clear M2 and M1
layers. We used these 2019 profiles to characterise how M2 and M1 peak
electron densities and altitudes change with SZA, LST sector - dawn vs
dusk - and season -- aphelion (Southern Autumn and Winter) vs perihelion
(Southern Spring and Summer). Therefore, we split these 219 profiles in four
groups: dawn aphelion, dawn perihelion, dusk aphelion, and dusk perihelion.
We find distinct differences between the different groups of data. The
biggest difference, both in peak densities and in peak altitudes, is found between dawn perihelion and aphelion, consistently with the hypothesis of a
more variable dawn ionosphere (Felici et al., 2022). For both the M1 and M2
layers, dawn perihelion is significantly higher in altitude and greater in density than dawn aphelion. For dusk aphelion and perihelion the difference is
usually less extreme. Dusk perihelion is higher in density than dusk aphelion,
but dusk aphelion is higher in altitude, particularly in the M1 layer.
Densities and altitudes of the M1 and M2 layers generally tend to be
higher at perihelion, expected since when Mars is closer to the Sun, we
expect higher solar flux to irradiate the atmosphere. However, densities and
altitudes are lower than what Fallows et al. (2015) found, at solar maximum.
Namely, corresponding to the solar minimum in the solar cycle, we find lower
peak densities and lower peak altitudes for both the M1 and M2 layers.
Finally, as SZA increases the M1 and M2 peaks get farther in altitude
from one another, yet closer in density".} 
Therefore, we utilised a novel approach in this study, and fully leveraged  the parameters estimated by Segale et al., [\href{https://arxiv.org/abs/2312.00734}{https://arxiv.org/abs/2312.00734}] to model the baseline electron density profiles from mere photochemical reactions, and subtract that baseline from our data. 
To so do, we estimated TEC, peak density, and peak altitude for two different sets of electron density profiles, namely: 
\begin{enumerate}
\item[{\bf SET A}] The series of measured ROSE electron density profile collected during time periods 1 ({\bf April 2021}) and 2 ({\bf June/July 2022}). An example of a measured electron density profile during period 2, affected by CIRs, ICMEs, and dust storms,  is reported in Figure \ref{fig_2}, in purple.
\item[{\bf SET B}] A series of modelled electron density profiles generated from eq. 2 in \cite{fallows2015a}, accounting from SZA dependence of the M2 and M1 peak densities and altitudes through their eq. 3 and 5. These equations are reported here as Equations \ref{eq1}, \ref{eq2}, \ref{eq3}, and \ref{eq4}.
The peak altitudes of the M1 and M2 layers follow 
\begin{equation}
\label{eq1}
z_{m} = z_{0} + L \times ln(\sec(SZA))
\end{equation}
where $m$ stands for either the M1 or the M2 peak, $z_{m}$ is the altitude of the peak, $z_{0}$ is the sub-solar altitude of the peak,  SZA is Solar Zenith Angle, and $L$ is the lengthscale which should coincide with the scale height of the neutral atmosphere, where the idealised photochemical theory applies. The peak densities of the M1 and M2 layers, instead, follow
\begin{equation}
\label{eq2}
N_{m} = N_{0}(\frac{1}{\sec(SZA)})^{k}
\end{equation}
 $N_{m}$ is the density of the peak, $N_{0}$ is the sub-solar density of the peak,  and $k$ is the fitted exponent.
For an idealised photochemical theory
\begin{equation}
\label{eq3}
N_{tot} = N_{m1} + N_{m2}
\end{equation}
Which leads to
\begin{equation}
\label{eq4}
N(z, sza) = \sum N_{m} exp(\frac{1}{2} (1 - \frac{z - z_{m}}{L_{m}} - exp(-\frac{z - z_{m}}{L_{m}})) )
\end{equation}
As values for $z_{0}, L_{m}, N_{m}$, and $k$ we utilised the fit parameters obtained \added{by} Segale et al.  [\href{https://arxiv.org/abs/2312.00734}{https://arxiv.org/abs/2312.00734}] and reported in their Tables  2 and 3, for appropriate SZA, and four different combination of Local Time Sector and L$_{S}$ range: dawn aphelion, dawn perihelion, dusk aphelion, and dusk perihelion. We so obtained $N_{m}$ and $z_{m}$, and, finally, $N$. Segale et al. [\href{https://arxiv.org/abs/2312.00734}{https://arxiv.org/abs/2312.00734}], in fact, not only identified the baseline for the variations in peak altitude and density of the M1 and M2 layers as a function of SZA from ROSE profiles of an undisturbed ionosphere at solar minimum  [see Segale et al., \href{https://arxiv.org/abs/2312.00734}{https://arxiv.org/abs/2312.00734}], but they separated these trends  into dawn (00:00 till 12:00 LST) and dusk (12:00 till 24:00 LST) sectors, and aphelion and perihelion season. As a reminder, aphelion and perihelion season are defined respectively as the intervals 0-161$^{\circ}$ and 341-360$^{\circ}$ L$_{S}$, and  161-341$^{\circ}$ L$_{S}$. Therefore, in this study, we were able to generate modelled electron density profiles for each of the profile in SET A, and account for their SZA, whether it was collected at dusk or at dawn, and if that was during perihelion or aphelion season.  An example of a so modelled electron density profile - of the merely photochemically produced undisturbed ionosphere - is reported in Figure \ref{fig_2}, in green: this is the modelled profile at the same SZA, local time sector, and season, as the corresponding measured profile (Figure \ref{fig_2}, in purple).
\end{enumerate}

Finally, we calculated the difference between TEC, peak density, and peak altitude of the profiles in {\bf SET A} and TEC, peak density, and peak altitude of the profiles in {\bf SET B} and obtained the three proxies 
$\Delta$TEC, $\Delta$(Peak Density), $\Delta$(Peak Altitude). In other words, $\Delta$TEC, $\Delta$(Peak Density), $\Delta$(Peak Altitude) constitute the deviation of TEC, peak density, and peak altitude induced by space weather and dust storms from the baseline photoproduced ionosphere. Besides being independent from SZA, LST sector, and season, these three proxies would be equal to 0 in undisturbed times for the ionosphere, namely times Segale et al., [\href{https://arxiv.org/abs/2312.00734}{https://arxiv.org/abs/2312.00734} ] explored in their study. We report the proxies and how they were derived in Figure \ref{fig_2}.

Since one of the two events was collected during 2021, approaching conjunction, to all the electron densities in all the profiles we attributed a maximum error of 1.5$\times$ 10$^{10}$ m$^{-3}$ \citep{Withers:2020um}, and the errors propagated accordingly. For the error associated to the TEC values, instead, we considered the difference between two different integration methods, one merely given by the manual integration of density values in altitude, one that utilises the Simpson method.

These three proxies from ROSE data were qualitatively compared  to the trends of MAVEN calibrated L2 in situ key parameters \href{https://pds-ppi.igpp.ucla.edu/search/view/?f=null&id=pds://PPI/maven.insitu.calibrated/document}{(on the PDS)} for SWEA (total flux of 1 - 500 eV and 500-1000 eV) \citep{mitchell2016},  
SEP (total flux for ions and electrons) \citep{larson2015}, and
dynamic pressure, from Solar Wind Ion Analyzer (SWIA) \citep{halekas2015}. In future work, we will conduct statistical studies on ICMEs impacting the ionosphere, and, to obtain meaningful quantitative information, utilising subsequent versions of the data for each instrument will be most appropriate, for it contains more accurate and recent quality flags. For the qualitative purpose of this study, the key parameters were a valuable resource instead.

To compare the different time series from in situ and remote instruments, and to identify the time delay between related events, we started from times of detection of the ROSE electron density profiles, and estimated backward time-lagged crossed correlation between ROSE derived proxies and Solar Wind (SW) particle parameters and irradiance. Time-lagged cross-correlation is a statistical technique used to measure the similarity  between two whole (not just minima and maxima) time-series while considering a time delay or lag between them. We considered lag = 0 hours/days between ROSE ($y$) and in situ ($x$), backward lag = 12hours/1day (MAVEN in-situ time series shifted -12 hours/-1 day from ROSE proxies), and backward lag = 24hours/2days (in-situ time series shifted -24 hours/-2 days from ROSE proxies). The cross-correlation function is computed for these three different time lags, so we could better visualise the similarity between the two time series at each lag. In other words, the cross-correlation function of x (in-situ solar wind particle parameters) and y (our ROSE derived proxies) is a matrix: the element at index k (the lag) in the resulting array is the correlation between {x[k], x[k+1], …, x[n]} and {y[0], y[1], …, y[m-k]}, where n and m are the lengths of x and y, respectively. A positive value of cross correlation indicates a positive correlation - in other words, similar trend in the two time series,  while a negative value suggests a negative correlation - or opposite movement of the two time series under examination. A value of zero means no correlation or similarity. In order to estimate the time-lagged crossed correlation, because the datasets did not have the same cadence, data had to be binned in 12 and 24 hours bins and the  mean in each bin of the MAVEN in-situ and remote datasets had to be estimated. 6 h bins led to a -0.1 $<$ correlation $<$ 0.1, larger than 24 h bins would have led to lose information on faster similarities.

The time period June/July 2022,  offer the unique opportunity of observing the behaviour of both the M1 and M2 layers in response to the A and B storm that were merging below the ionosphere at this time: we investigate whether of not the M1 and M2 layers are lofted together and if the response of the ionosphere to CIR and/or ICME changes if dust storms are lofting the ionosphere from below.

We report and discuss the results of this methodology in the next Section (\ref{results}).

\section{Results and discussion}
\label{results}
\subsection{April 2021: CIR and ICME impact at aphelion}
\label{april}

\begin{figure}
	\centering 
	\includegraphics[width=0.7\textwidth]{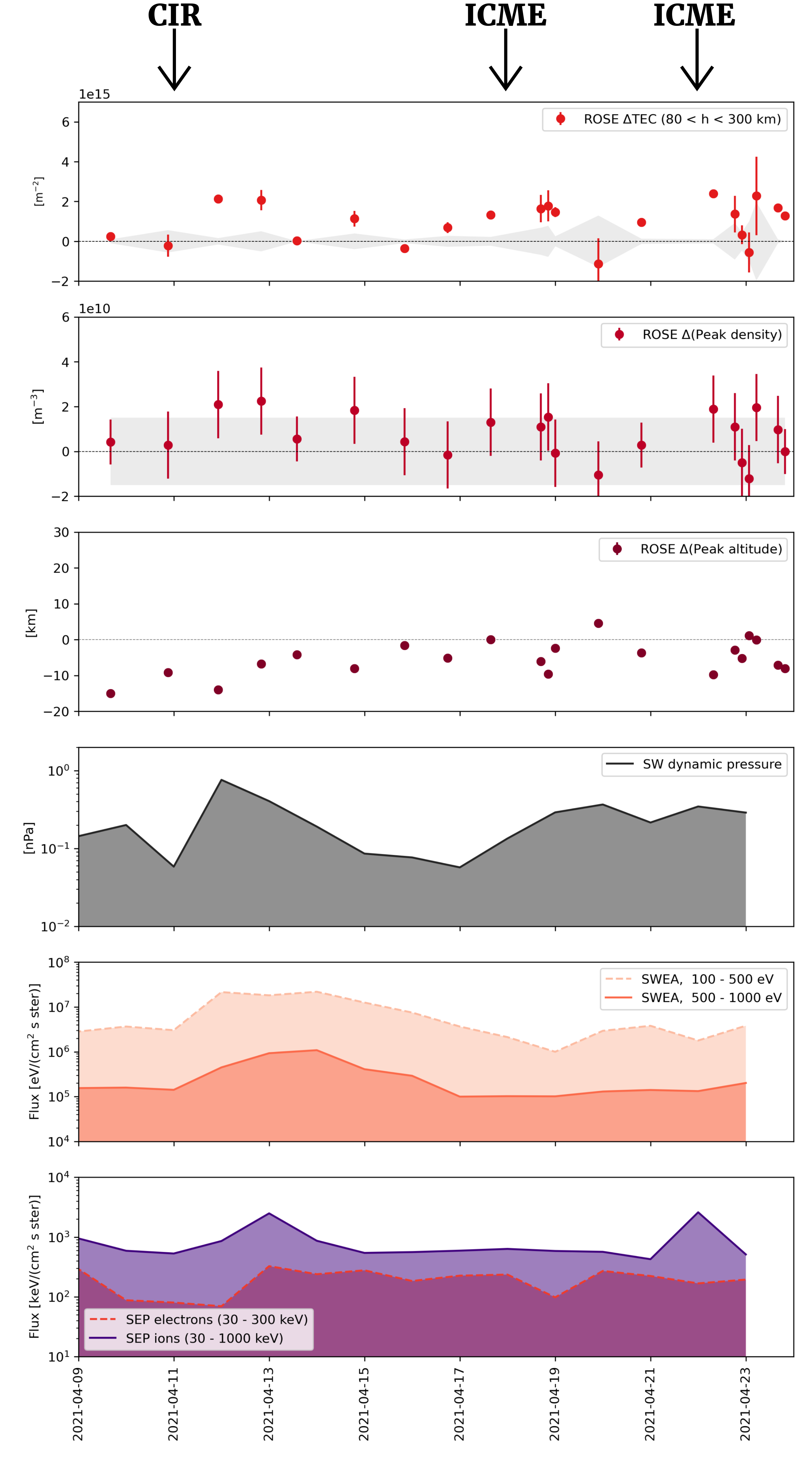}	
	\caption{April 2021: From top to bottom, time series  of ROSE $\Delta$TEC, ROSE $\Delta$(Peak density), $\Delta$(Peak altitude); solar wind dynamic pressure derived from MAVEN SWIA; SWEA electron flux; SEP ions and electron fluxes. The first three proxies, all derived from ROSE data, are independent from L$_{S}$, SZA, and LST, therefore we only report their trend\added{, and we indicate with grey shadows the error estimated as described in Section \ref{methods} for each proxy}. At the top of the plot we indicate arrival times of the CIR and ICMEs in this time interval as predicted by the ENLIL model. } 
	\label{fig_3}%
\end{figure}

As mentioned in Section \ref{methods}, this time interval is covered by 21 ROSE electron density profiles collected between April 9th and April 23rd 2021. These profiles span a range in SZA between $\simeq$ 71 and 94$^{\circ}$ and were collected close to aphelion. During this time period, based on the ENLIL simulation results available on the \href{https://kauai.ccmc.gsfc.nasa.gov/DONKI/search/}{Space Weather Database Of Notifications, Knowledge, Information (DONKI) catalog}, in April 2021, the first increase in particle flux  might be due to a CIR impacting Mars around April 11th, the second to a glancing blow from an ICME impacting Mars on April 18th, the third to a direct ICME impact on Mars on April 22nd (Figure \ref{fig_a1}, and we report the timings for these modelled impacts at the top on Figure \ref{fig_3}).
In Figure \ref{fig_3}, we report the three proxies estimated from ROSE parameters, namely $\Delta$TEC, $\Delta$(Peak Density), and $\Delta$(Peak Altitude) (we want to remind the reader that these proxies are independent from SZA, LST sector, L$_{S}$ sector);  the Solar Wind (SW) dynamic pressure estimated through the SWIA instrument on MAVEN; the SWEA flux measurements the particle energy ranges 100-500 and 500-1000 eV; and the SEP meaurements for ions (energy 30-1000 keV), and SEP electrons (30-300 keV). Follow SWEA and SEP in the panels below.  

The first proxy shown in Figure \ref{fig_3}, top panel, is the $\Delta$TEC: many of the profiles show a larger TEC than the expected values. For times around April 11th, this corresponds to a larger peak density as well (Figure \ref{fig_3}, second panel), anyway consistent with 0 within error.
$\Delta$TEC reaches values of $\simeq$ 2.5 $\times 10^{15}$ m$^{-2}$ on April 22nd, and $\simeq$ 2 $\times 10^{15}$ m$^{-2}$ on April 19th without any striking enhancement in $\Delta$(Peak Density) or $\Delta$(Peak Altitude). Therefore, the additional ionisation enhancing the $\Delta$TEC must be occurring altitudes different from that of the main peak, as precipitating particles and EUV radiation are different phenomena. In fact, \cite{Nakamura:2022aa}  estimated that 1 keV electrons can ionise CO$_2$ down to 120 km altitude, therefore particles detected by the SWEA instrument should be responsible only for ionisation - besides the usual photochemical ionisation - at and above the main ionospheric peak. \cite{Nakamura:2022aa} also estimated that electrons above 50 keV and protons with energy 500 keV  - where both these energies can be measured by MAVEN SEP - can ionise CO$_2$ deeper in the atmosphere, down to 80 km altitude, with CO$_{2}^{+}$ production rates 2-3 order of magnitudes larger than that produced by 1 keV electrons.

\begin{figure}
	\centering 
	\includegraphics[width=\textwidth]{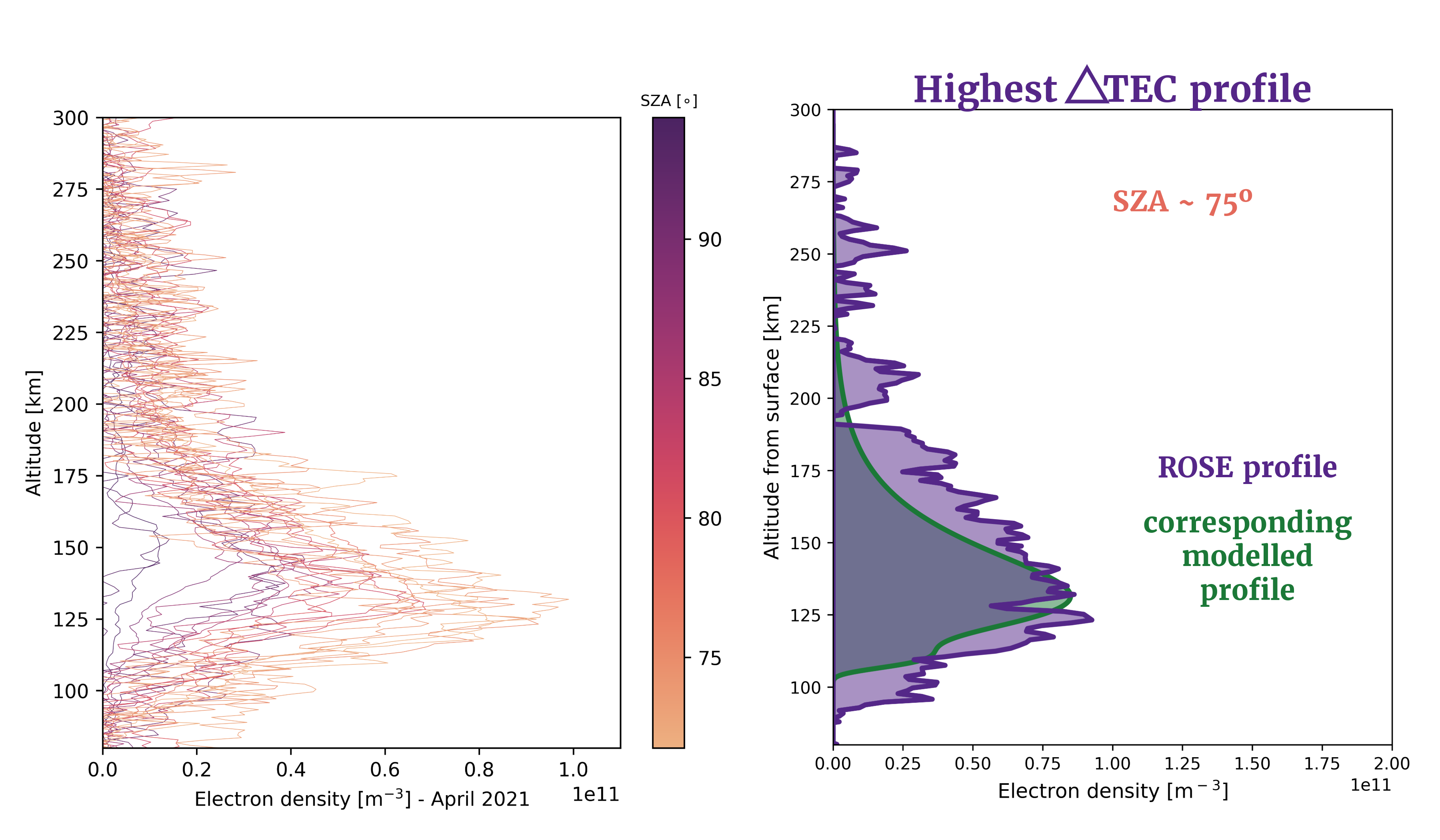}	
	\caption{April 2021: (left) ROSE electron density profiles in the time interval covered in Figure \ref{fig_3}, with colourscale indicating values SZA: these are the original electron density profiles belonging to {\bf SET A} described in Section \ref{methods}. (right) ROSE electron density profile that see the highest increase in TEC (purple) compared to the baseline profiles (green).} 
	\label{fig_4n}%
\end{figure}

We report the original ROSE electron densities profiles for this time interval in Figure \ref{fig_4n}, left, with colourscale representing the SZA of the profiles: these are really disturbed and noisy electron density profiles, possibly due to the vicinity with Mars solar conjunction \cite{Withers:2020um}. In general, the expected behaviour consistent with an idealised photochemical theory - smaller densities with SZA, higher peak altitudes with SZA - can be appreciated, however with the presence of bulkier plasma content at lower altitudes (below 100 km) for a few of these profiles: MAVEN SEP shows an increase in flux of the Solar Energetic Particles (SEPs), able to produce plasma layers below $\sim$ 90 km altitude \citep{SanchezCano:2019aa, Peter:2021aa,Lester:2022aa, Harada:2023aa}. Additionally, these profiles tend to display a rather extended and spread in altitude ionosphere and,  wider-than-the-baseline ionospheric layers (see  Figure \ref{fig_4n}, right, comparison between the measured profile and the modelled), indicating more ionisation than that only due to photoionisation, possibly due to an interplay of increased compression and precipitating particles from the SW: as the high pressure causes the solar wind to get into the ionosphere \citep{Dubinin:2009aa}, solar particles at various energies can penetrate down to various altitudes getting not only down to 80 km \citep{Ulusen:2012aa, Sanchez-Cano:2017aa}, but to the ground \citep{Zeitlin:2018aa,Ehresmann:2018aa}.
Another rather striking feature shown by these proxies is the suppressed $\Delta$(Peak Altitude) throughout this time period (see Figure \ref{fig_3}, third panel), by more than 15 km in the beginning of this time interval, possibly due to a response to the increased dynamic pressure, which was observed in previous studies \citep[][]{Sanchez-Cano:2017aa}. The $\Delta$(Peak Altitude) fluctuates between $\simeq$ - 10 and 8 km after April 11th.

\begin{figure}
	\centering 
	\includegraphics[width=\textwidth]{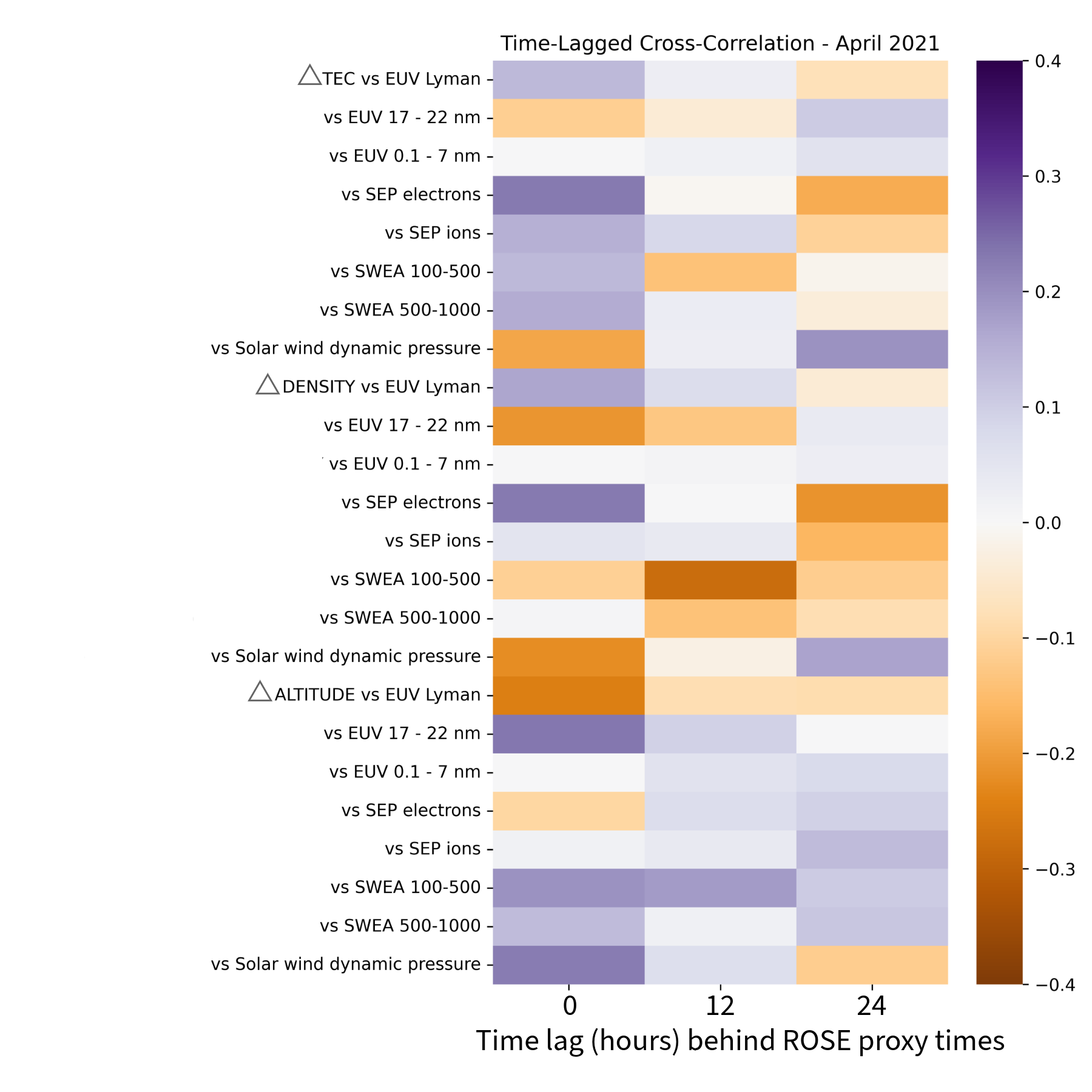}	
	\caption{April 2021: Backward time lagged crossed correlation matrix, displayed as a heatmap where, on the y axis, we represent days of lag between the \added{semi-}daily averaged solar particles parameters measured by MAVEN in-situ instruments, and the x axis reports the various quantities between which the backward time lagged cross correlation was estimated. The colourscale represent correlation values} 
	\label{fig_5}%
\end{figure}

To gain more focus on the temporal correlation, attempt to untangle the many factors garbled together, and not to attribute causation, we estimated the time-lagged cross correlation between the solar wind particle parameters in Figure \ref{fig_3}, the solar irradiance measured with MAVEN Extreme Ultraviolet Monitor (EUVM) \citep{Eparvier:2015wc}, and the ROSE proxies, and we show the time-lagged crossed correlation matrix through the heat map in Figure \ref{fig_5}: each element, or correlation value, in the matrix is represented in a cell in the map. There is no strong correlation or anti-correlation between time-series, we are anyway below $|0.5|$ values, however we can use this map as support information to accompany Figure \ref{fig_3}: a high clear correlation between two time-series cannot be expected in this case, because there are multiple solar wind parameters modulating the values each single proxy. In other words, solar wind high energy and thermal particles, fields, solar irradiance, these are all affecting the ionosphere: we would not expect to see here a cross correlation $=$ 1 between two time series. However, this matrix can help us see which time series are the most similar to each other more intuitively than Figure \ref{fig_3}. It is also important to remind the reader that MAVEN is not upstream, therefore the timing of the in situ parameters may not reflect perfectly the time at which the atmosphere is impacted. 

The most similar times-series of parameters and in situ data binned in averaged 12 hours bins, are $\Delta$TEC to SEP electrons time-series, $\Delta$(Peak Density) and to SEP electrons, $\Delta$(Peak Altitude) to EUV 17-22 nm wavelengths and to the solar wind dynamic pressure, and with lag = 0. The most \replaced{anti-correlated}{dissimilar} time-series are the  $\Delta$(Peak Altitude) to EUV Lyman $\alpha$ with lag = 0, and the $\Delta$(Peak Density) to SWEA 100 - 500 eV: photons with wavelenght $\sim$ 121.6 nm or electrons with energy between 100 - 500 eV might not have enough energy to penetrate down to the main peak. However, electrons with energy between 100 - 500 eV can ionise higher altitudes, while SEP electrons can ionise down to 80 km (the lower limit of our integration).

This might suggest that the ionisation during CIRs and ICME impacts is more spread in altitude and relies on mechanisms additional to photoionisation, differently to what happens in a quiet ionosphere mainly generated by photochemistry, consistently with what \citep{Ulusen:2012aa, SanchezCano:2019aa} found.

\subsection{June - July 2022: CIR and ICME impact during dust season and a dust storm}

\begin{figure}
	\centering 
	\includegraphics[width=0.7\textwidth]{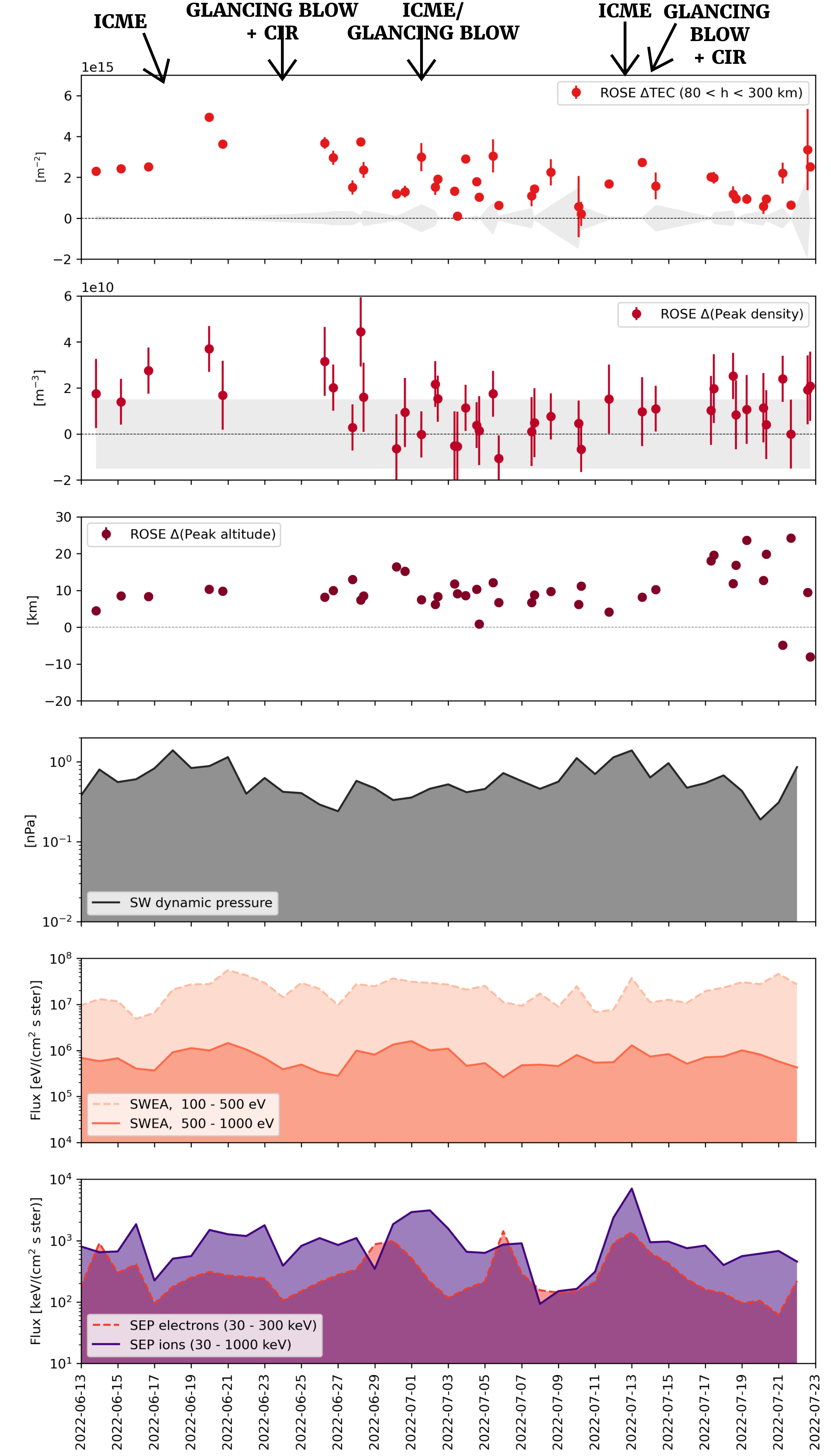}	
	\caption{June - July 2022: as Figure \ref{fig_3}.} 
	\label{fig_6}%
\end{figure}

This time interval included \replaced{41}{42} electron density profiles collected between June 13th and July 22th 2022, a SZA range between $\simeq$ 47 and 95$^\circ$. These profiles were collected not only during dust season, but also during  dust storms (see Section \ref{introduction}, and Figure \ref{fig_1}). In fact, one A storm and one B storm were taking place continuously,  overlapping, in the entire time interval under examination (see Section \ref{methods}). 
Besides the dust storms lofting the ionosphere from below, in this time interval the ionosphere is also affected by solar events from above: based on the ENLIL simulation results, there may have been an ICME impact at Mars on June  18th, a glancing blow from an ICME combined with a CIR on June 24th, a glancing blow from an ICME on July 2nd, an ICME on July 13th, and a glancing blow from an ICME combined to a CIR on July 14th (see Figure \ref{fig_a2}, and arrows on top of Figure \ref{fig_6}). MAVEN is upstream of Mars continuously only after July 6th (see Figure \ref{fig_a3}).

In Figure \ref{fig_6}, we report the three proxies derived from ROSE data, namely $\Delta$TEC, $\Delta$(Peak Density), and $\Delta$(Peak Altitude), then the SW dynamic pressure, then SWEA measurements, and SEP measurements. The SW parameters in this time interval are stronger than during April 2021 (see Figure \ref{fig_3}).
Definitely, the ionosphere in this time period is subject to significant stress from above and below, and this is reflected in the trends of our three proxies:
$\Delta$TEC starts from $\simeq$ 2 $\times 10^{15}$ m$^{-2}$ (one of the maximi values in April 2021)
 in this time interval and reaches $\simeq$ 5 $\times 10^{15}$ m$^{-2}$
 on June 20th, fluctuating between $\simeq$ 4 $\times 10^{15}$ m$^{-2}$ and 0 throughout the rest of this time interval. Particularly prominent, the increase to $\simeq$ 4 $\times 10^{15}$ m$^{-2}$ on July 22nd, in correspondence to an increase in upstream $\alpha$ density which can be appreciated in Figure \ref{fig_a2}.

$\Delta$(Peak Density) is quite elevated from the beginning of the interval, until June 26th, to then proceed to be consistent with 0, or slightly above it, until the end of the interval. 

What is dramatically different in June/July 2022 compared to April 2021 is the $\Delta$(Peak Altitude): it fluctuates consistently around values 5-16 km, until July 17th, when it reaches an extraordinary $\sim 25$ km, and negative values in two instances. These extraordinary 25 km, and other $\Delta$(Peak Altitude) $\sim$ 20 km seem to correspond to a depressed solar wind dynamic pressure. 
 The  $\Delta$(Peak Altitude) falling abruptly below 0 might correspond to two profiles collected at SZA $\simeq$ 94 - 95$^{\circ}$, possibly reflecting the condition of the nightside ionosphere, not the photochemical dayside ionosphere.

Dust storms can increase the peak altitude up to $\sim$ 20 km \citep[][and references therein]{felici2020}, yet the $\Delta$(Peak altitude) ranges between 0 and 16 km, staying on average $\sim$10 km until July 17th, at the end of the time interval examined, when it reaches $>20$ km values. These higher values of $\Delta$(Peak altitude) correspond to times at which the solar wind dynamic pressure was lower than the rest of the interval.

Whether or not this specific dust storm would have raised the peak altitudes more if the SW dynamic pressure (see Figure \ref{fig_6}) were not as strong in this time interval, is hard to determine with certainty. However, based on the suppressed trend of $\Delta$(Peak Altitude) - negative, on average -  during merely CIRs and ICME impacts (see Section \ref{april}), and previous \added{observations collected during dust storms}  \cite[e.g.,][and references therein]{Girazian:2019, felici2020}, previously observed suppression of the peak altitude corresponding to higher solar wind dynamic pressure \citep{Sanchez-Cano:2017aa}, we suggest that it might have. To be noted is that part of the increase the peak altitude up to $\sim$ 20 km \citep[e.g.,][and references theirein]{Girazian:2019, felici2020} observed might have been due to mere perihelion effects, or dawn vs dusk effects (see Segale et al., \href{https://arxiv.org/abs/2312.00734}{https://arxiv.org/abs/2312.00734}) not considered in those studies. However, the fact that the high SW dynamic pressure might suppress the raise in peak altitude of the thermosphere during a dust storm could also help explain why \cite{withers2013} found a mere $\simeq$ 5 km increase in peak altitude in the MGS data: during solar maximum, solar activity is higher, therefore there might have been factors from above restraining the loft of the atmosphere induced, from below, by dust.

\begin{figure}
	\centering 
	\includegraphics[width=\textwidth]{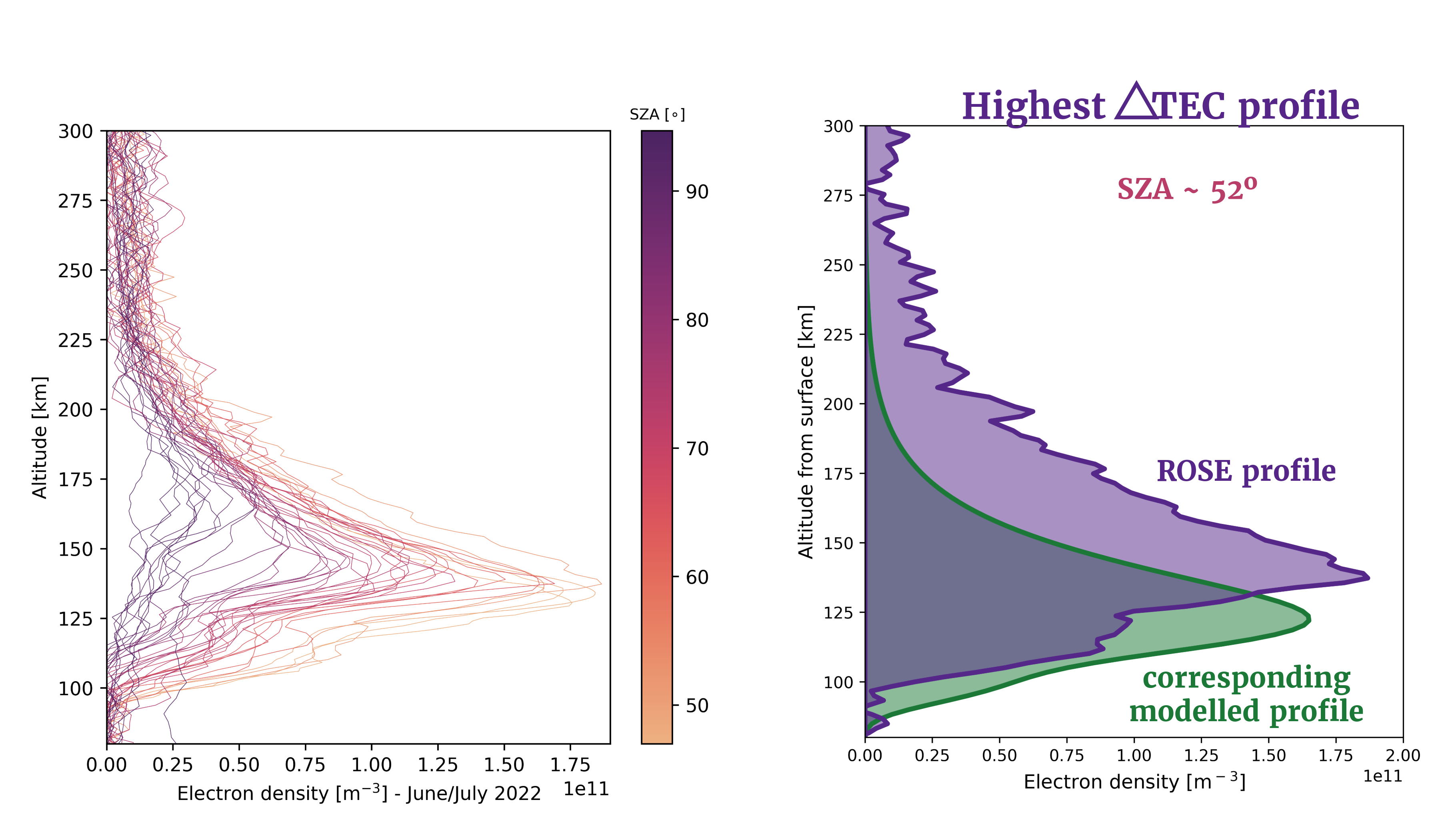}	
	\caption{June - July 2022: as Figure \ref{fig_4n}, for this time interval. } 
	\label{fig_7}%
\end{figure}

We report in Figure \ref{fig_7} the series of ROSE electron density profiles for this time period, with colourscale representing SZA (right). We can notice how these profiles are way less noisy than the ones in Figure \ref{fig_4n}, how pronounced the M1 layer is in some of these profiles (less expected outside of solar maximum \cite{Withers:2023aa}), and how extended to higher and lower altitudes the ionosphere is. The latter point can be further appreciated in Figure \ref{fig_7}, right, where we report\deleted{, as in \ref{fig_4n},} the ROSE \replaced{profile}{profiles} that lead to the highest $\Delta$TEC (in purple), with the corresponding modelled \replaced{profile}{profiles} (green). 

\begin{figure}
	\centering 
	\includegraphics[width=\textwidth]{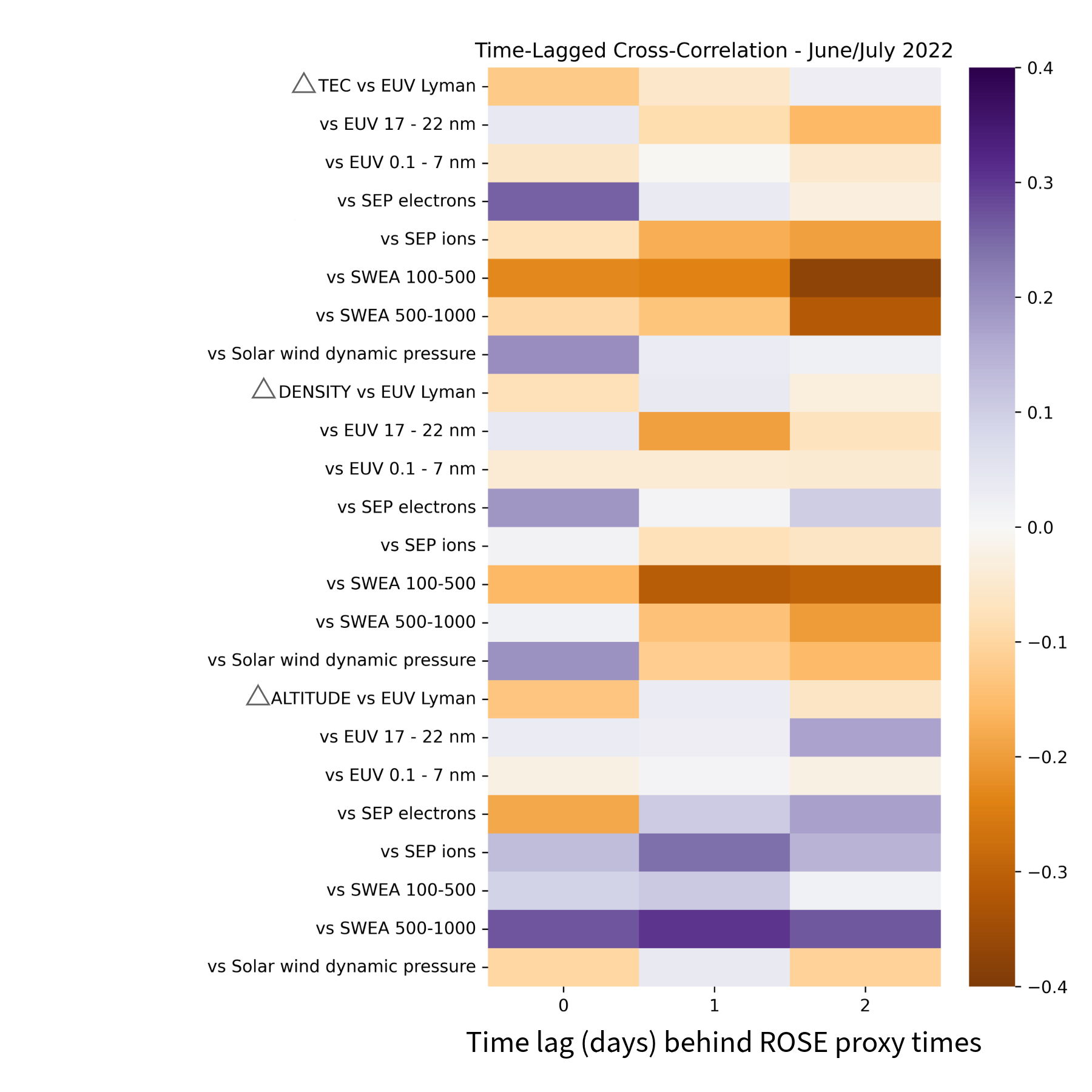}	
	\caption{June - July 2022: as Figure \ref{fig_5}, for this time interval. } 
	\label{fig_8}%
\end{figure}

As before, we use the time-lagged crossed correlation matrix, displayed in Figure \ref{fig_8} through a heatmap, to help us compare \added{24h bin averaged} time-series trends, with similar caveats to the previous case (e.g. low correlation values). \added{In this case, we had to adopt 24h bin size because the 12h bin size, not shown here, resulted in no correlation at all.} In this case, $\Delta$TEC time-series is most similar to the SEP electrons time-series with lag = 0.
Similarly to what we saw in Figure \ref{fig_5} for April 2021, $\Delta$(Peak density) seems  still \added{loosely} anti-correlated to lower energy SWEA electrons fluxes (which have energy comparable to the 10 nm photons, though with different cross section). The $\Delta$(Peak altitude) is \replaced{more}{most} correlated with the higher energy SWEA fluxes than for April 2021 for time lag = 0,1,2.

\begin{figure}
	\centering 
	\includegraphics[width=\textwidth]{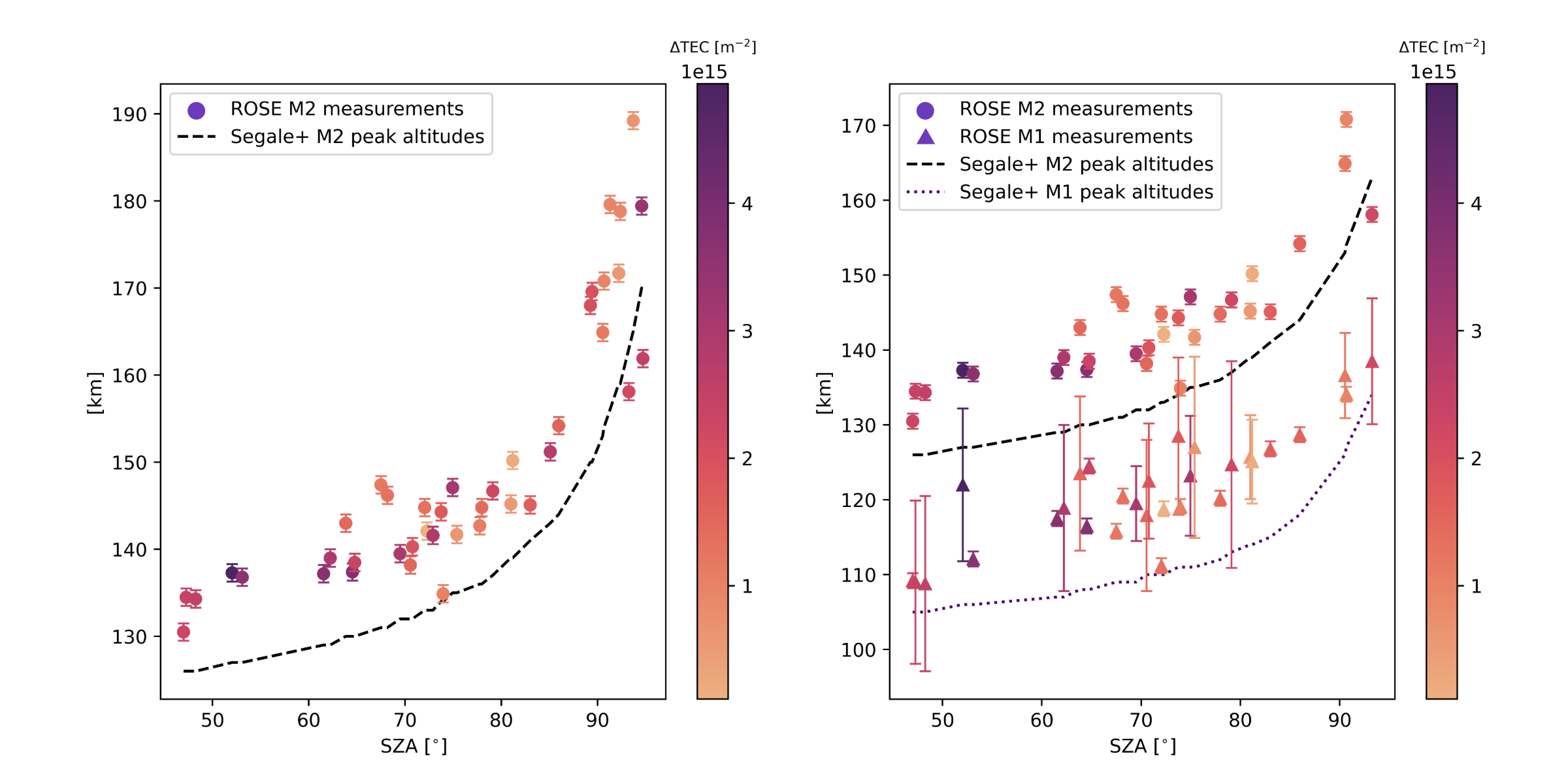}	
	\caption{June - July 2022: (Left) All (41) M2 peak altitudes (circles)  for the time interval in the timeline in Figure \ref{fig_8}, compared to baseline values from their LST time sector and L$_{S}$ around perihelion reported by Segale et al. [\href{https://arxiv.org/abs/2312.00734}{https://arxiv.org/abs/2312.00734} \added{(Right)} M2 peak altitudes (circles) and M1 peak altitudes (triangles) for the time interval in the timeline in Figure \ref{fig_8}, \added{for the electron density profiles that show both M2 and M1 layers (30), } compared to baseline values from their LST time sector and L$_{S}$ around perihelion reported by Segale et al. [\href{https://arxiv.org/abs/2312.00734}{https://arxiv.org/abs/2312.00734}] for both layers. } 
	\label{fig_9}%
\end{figure}

\replaced{30 of the 41}{31 of the 42} profiles in this time interval displayed, besides the ever-present-in-the-dayside M2 layer, also the M1 layer (see Figure \ref{fig_7}), allowing us to observe how both the M1 and M2 layers, therefore different ionospheric altitudes, respond to a dust storm, while compressed and percolated by solar particles and fields. We show in Figure \ref{fig_9}\added{(left) the 41 peak altitudes as a function of SZA and $\Delta$TEC, and in Figure \ref{fig_9}(right)} the altitude of the peak of both layers as a function of SZA, and also the baseline values obtained by Segale et al., [\href{https://arxiv.org/abs/2312.00734}{https://arxiv.org/abs/2312.00734}] for the ionosphere at dawn and dusk at perihelion: both M2 and M1 peak altitudes are above the baseline expected (but one \added{profile that has a M1 layer as well as the M2}, the last in this set and two in total, see Figure \ref{fig_6}), indicating that the thermosphere was \added{generally} lofted by the dust storm underneath. Moreover, \added{for profiles that display both M1 and M2 layers,} the M2 peak altitudes display a distance from their baseline  (also reported in Figure \ref{fig_6} bottom panel, as $\Delta$(Peak Altitude) ) between -4.9 (only one point has negative value) and 16.8 km, and the M1 between -0.5 and 17.5 km, where these two ranges are consistent within error. This suggests that the M2 and M1 layers are not affected differently by a dust storm: the thermosphere, on average, lofts as a unit.

\begin{figure}
	\centering 
	\includegraphics[width=\textwidth]{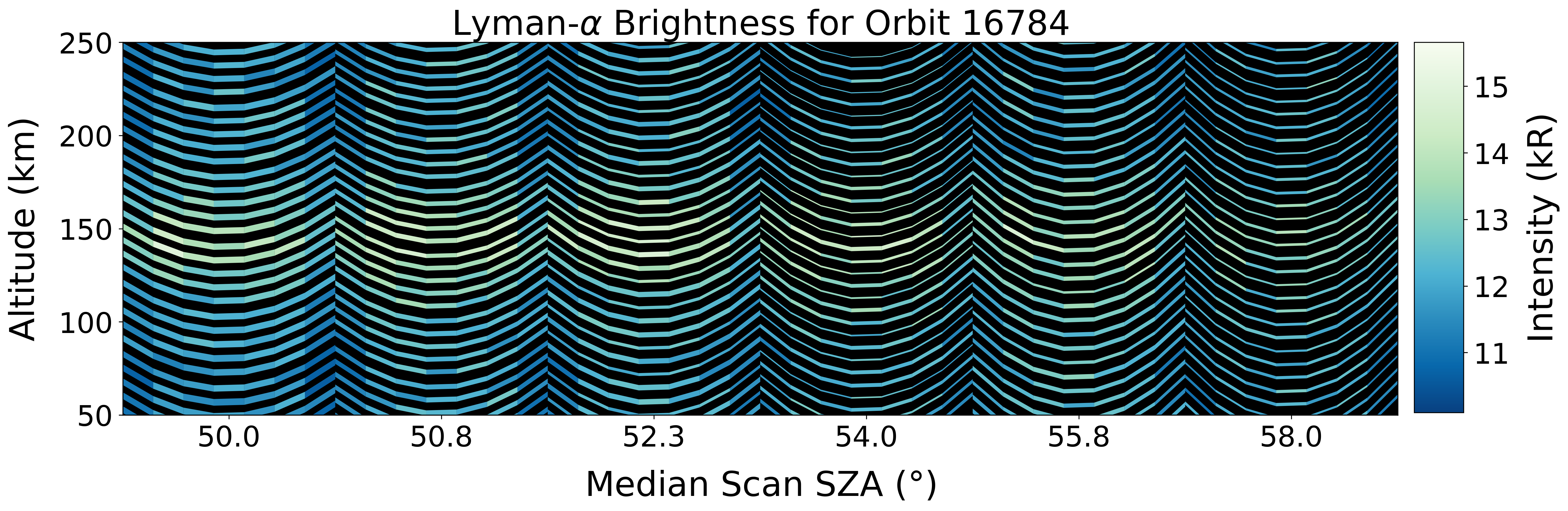}	
	\caption{A synthetic image format of IUVS Ly$-\alpha$ intensity (see text for details), showing an enhancement in Ly$-\alpha$ above the background coronal H due to proton aurora activity during this orbit. This IUVS observation was taken during the outlimb portion of the MAVEN orbit (Orbit n.16784 on 13 July 202203:34 UTC), and occurs during the same time period as the ROSE observations shown in Figure \ref{fig_6}, corresponding to a local maximum in $\Delta$TEC, demonstrating the widespread effect of dust activity and extreme solar activity on the dayside of the planet.} 
	\label{fig_10}%
\end{figure}

To add to the global picture of the ionosphere when triggered by solar drivers, aurora enhancement during an ICME was previously detected by \cite{Jakosky:2015CME}. Proton aurora at Mars commonly occur across the entire dayside of the planet around the southern summer solstice \citep{Hughes:2019aa}. This time period overlaps with the Martian dust storm season when atmospheric dust and temperatures reach an annual high, leading to increased atmospheric H escape rates and significant inflation of the hydrogen (H) corona beyond the bow shock, and in turn, more frequent and bright proton aurora activity during this time  \citep{Hughes:2019aa, Chaffin:2021aa}. The Imaging Ultraviolet Spectrograph (IUVS) instrument \citep{McClintock:2015aa} on board the MAVEN spacecraft observed multiple notable proton aurora events during the months of June and July of 2022 (e.g., Figure \ref{fig_10}). These events are observed in the IUVS limb scan data as enhancements in the H Lyman-$\alpha$ (Ly-$\alpha$) emission (121.6 nm) above the background coronal H brightness. Figure \ref{fig_10} shows an example proton aurora detection during the time period of this study (Orbit 16784 on 13 July 202203:34 UTC). Figure \ref{fig_10} presents a synthetic image format of the un-binned Ly-$\alpha$ limb scan data acquired during the outlimb portion of the MAVEN orbit; i.e., each of the six outlimb IUVS limb scans (horizontally aligned panels) displays the Ly-$\alpha$ intensity for each of the 21 IUVS mirror integrations (vertical) and seven spatial bins within each scan (diagonal/horizontal rectangular bins) (e.g., similar to Figure 2 from \cite{Deighan:2018aa}). Note that although all of the limb scans are displayed side-by-side in the figure, there is actually a spatial separation between each scan. The highest intensity bins in the $\sim$110-150km altitude range correspond with a Ly-$\alpha$ enhancement above the background coronal H that is associated with proton aurora activity. Numerous proton aurora events observed during this time period correspond with increases in the ROSE TEC, peak density, and/or peak altitude, demonstrating the widespread and multifaceted impact of dust activity and extreme solar activity on the dayside of  the planet.

\section{Summary and conclusions}
\label{conclusion}
We utilised the study conducted by Segale et al., [\href{https://arxiv.org/abs/2312.00734}{https://arxiv.org/abs/2312.00734}] as a baseline for the \added{photochemical} Martian ionosphere \replaced{through solar minimum leading to solar maximum}{at solar minimum}, during dust season, or perihelion season, and aphelion season, and at dawn and dusk, to define three proxies from ROSE electron density profiles - $\Delta$TEC, $\Delta$(Peak Density), and $\Delta$(Peak Altitude). The purpose of defining these three proxies is to estimate the deviations from a photochemical ionosphere when the ionosphere is subject to external drivers, such as CIRs, ICMEs, and dust storms. In fact, we used these three proxies to study the response to solar events that hit Mars at aphelion and during dust storms.
While there are too many different factors at play to be able to make definitive statements about causality from the analysis of only two time periods, this study suggests that:
\begin{enumerate}
\item an increase in SEP particles flux (seen with MAVEN SEP) can correspond to an increase in ionisation down to 80 km altitudes, with a short ($<$ 1 day) thermospheric response, increasing the TEC between 80 and 300 km altitude up to $\simeq$ 2.5 $\times 10^{15}$ m$^{-2}$ in April 2021 and up to $\simeq$ 5 $\times 10^{15}$ m$^{-2}$ in June/July 2022. An increase in electron fluxes detected with SWEA can correspond in time to an increase in the altitude at which highest electron density is measured.
\item in June/July 2022, an A storm and a B storm were ongoing and merging under the ionosphere throughout the time period we examined. From 31 profiles that showed both the M2 and M1 layer, we observe that, on average, M1 and M2 peak altitudes raise the same amount, suggesting that the thermosphere might loft as a unit. Whether or not high SW dynamic pressure suppresses the raise of the peak altitudes is harder to determine, however, comparisons between the $\Delta$(Peak Altitude) in April 2021 and the $\Delta$(Peak Altitude) in June/July 2022, and with results from previous studies suggest that that might be the case.
\item this time period corresponds to the detection of several proton aurora events, detected with MAVEN IUVS, of which the brightness modulates as well as ROSE TEC, peak density, and/or peak altitude do, underlining how complex and multifaceted the impact of dust activity and extreme solar activity on the Martian ionosphere can be.
\end{enumerate}

The approaching of solar maximum, and of dust season during solar maximum, will provide us with more opportunities to conduct systematic and statistical studies of the impact of solar events on the Martian ionosphere with or without dust storm lofting the thermosphere\added{, coupled with future proton aurora studies}.

\section*{Acknowledgements}
 This work was supported by NASA under award number NNH1OCCO4C, LASP subcontract 9500306435. MAVEN key parameters data bundle available on \href{https://pds-ppi.igpp.ucla.edu/search/view/?f=yes\&id=pds://PPI/maven.insitu.calibrated}{https://pds-ppi.igpp.ucla.edu/search/view/?f=yes\&id=pds://PPI/maven.insitu.calibrated}. Simulation results have been provided by the Community Coordinated Modeling Center (CCMC) at Goddard Space Flight Center through their publicly available simulation services (\url{https://ccmc.gsfc.nasa.gov}). The ENLIL Model was developed by Dusan Odstrcil, at George Mason University.

 \newpage
 \appendix
 \section*{Appendix A}
 \label{appa}
 \counterwithin{figure}{section}
 \begin{figure}
 	\centering 
 	\includegraphics[width=\textwidth]{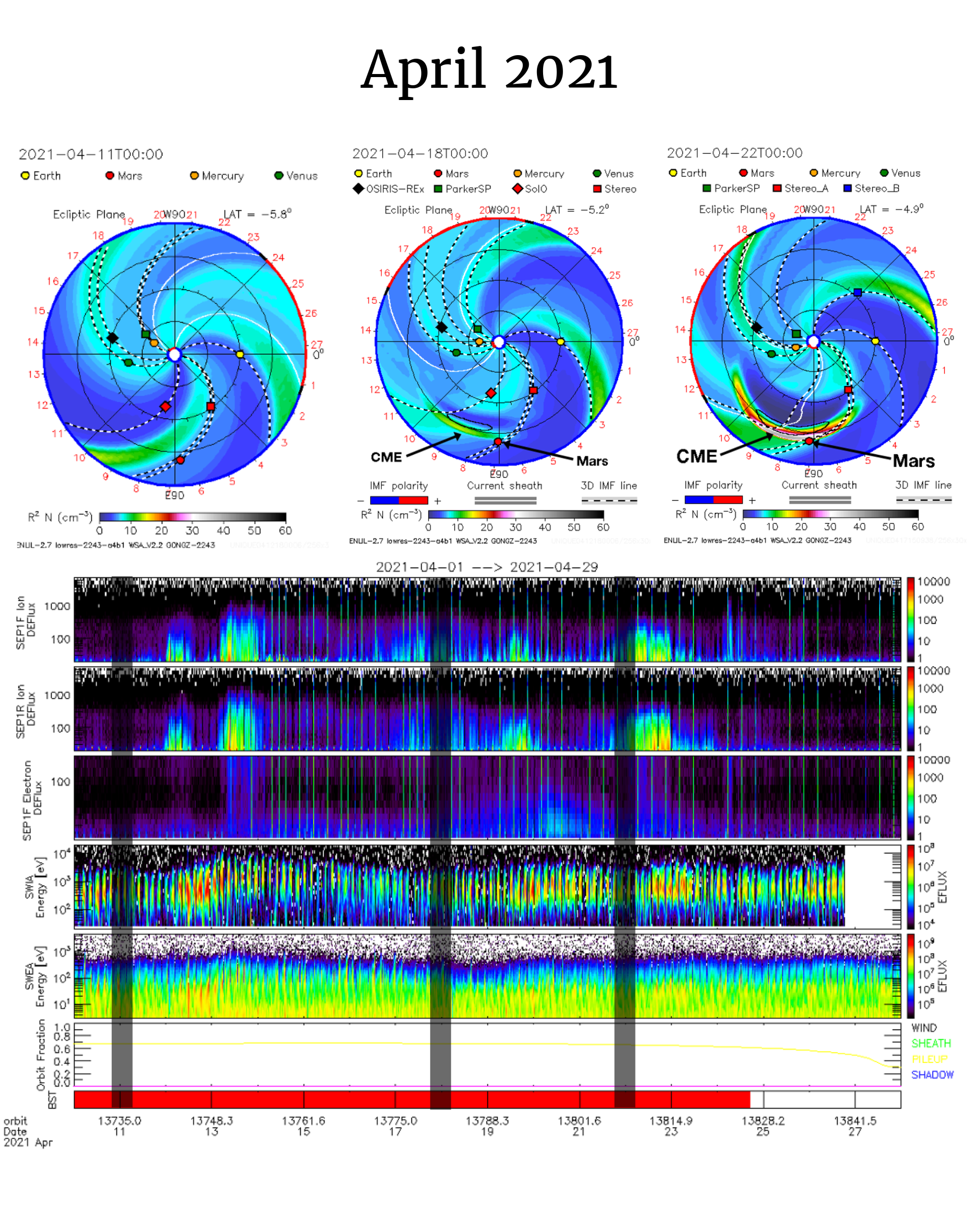}	
 	\caption{April 2021: heliospheric conditions results of ENLIL runs for this time period (top); MAVEN in situ SEP and SWEA measurements. \added{We highlighted with black shadows predicted solar events arrival time}.} 
 	\label{fig_a1}%
 \end{figure}

 \newpage

 \begin{figure}
 	\centering 
 	\includegraphics[width=1\textwidth]{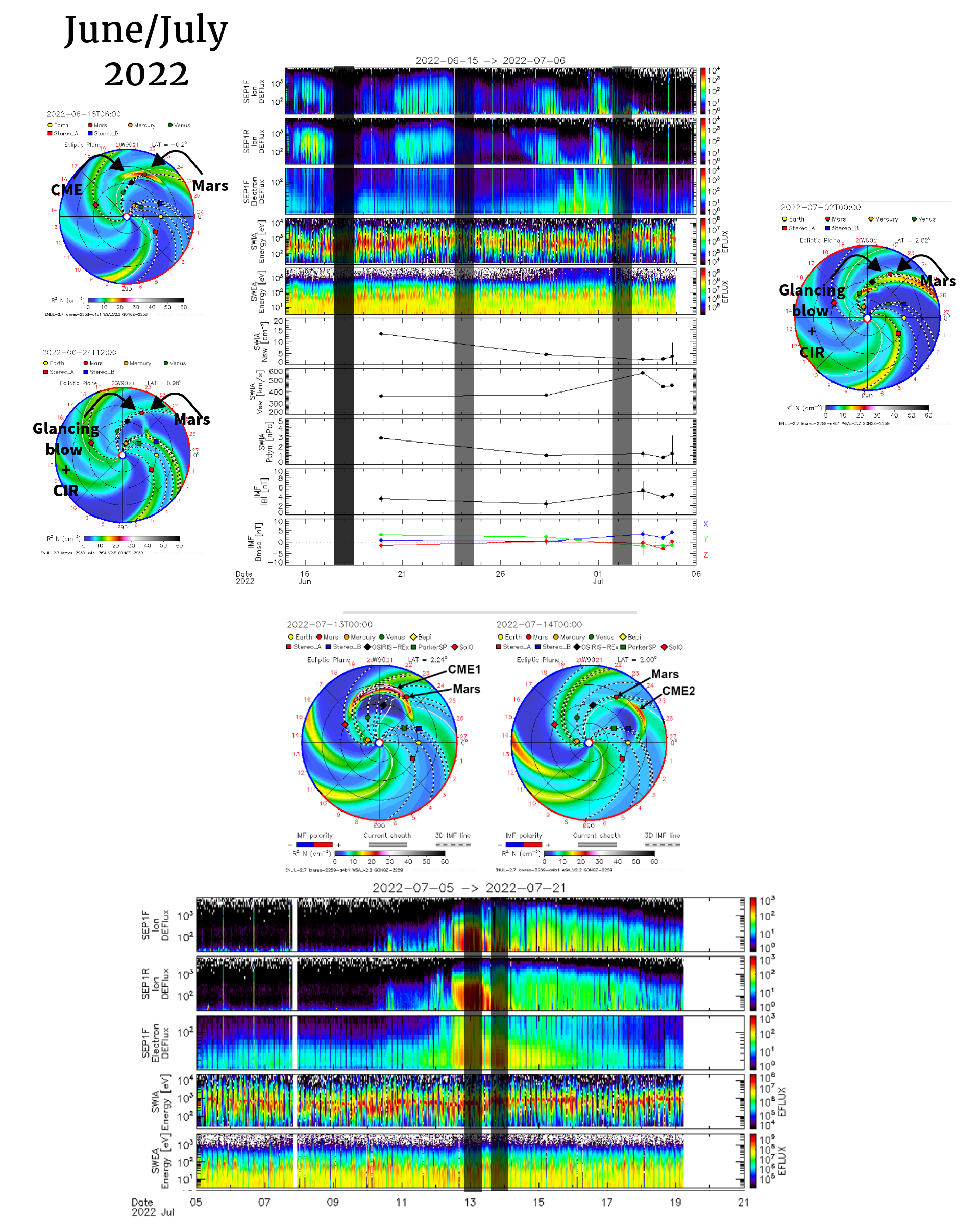}	
 	\caption{June/July 2022: as Figure \ref{fig_a1}, but for this time period. } 
 	\label{fig_a2}%
 \end{figure}
 \newpage
 \begin{figure}
 	\centering 
 	\includegraphics[width=0.68\textwidth]{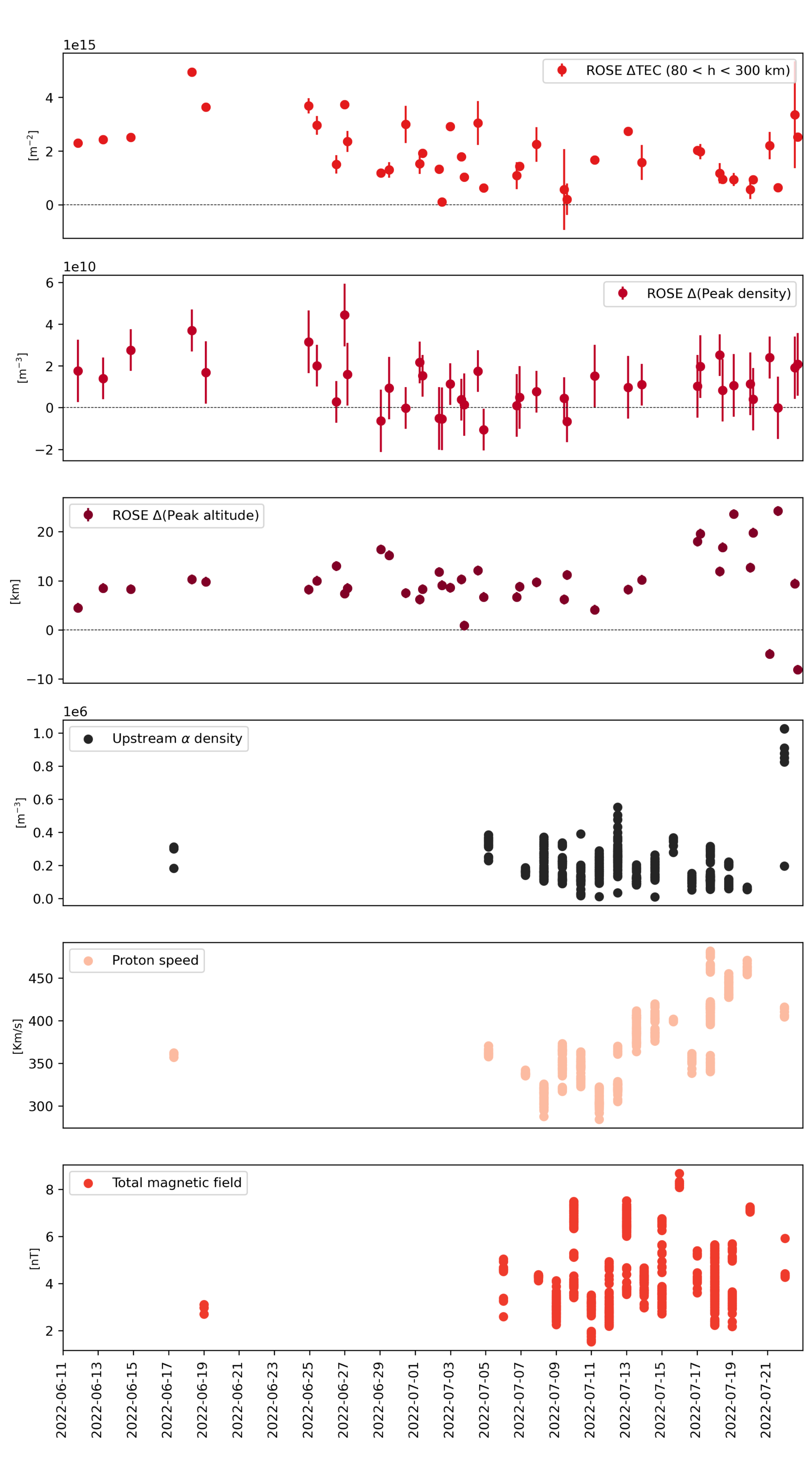}	
 	\caption{June/July 2022: as Figure \ref{fig_3} for the top three panels. The bottom three panels show MAVEN upstream high resolution data only for $\alpha$ density, proton speed, and SW total magnetic field \citep{halekas2015, Halekas:2017aa, Halekas:2015aa}, available at \url{https://homepage.physics.uiowa.edu/~jhalekas/drivers.html}.} 
 	\label{fig_a3}%
 \end{figure}
 \newpage
  \clearpage
\bibliographystyle{elsarticle-harv} 

\end{document}